\documentclass[conference,letterpaper]{IEEEtran}
\usepackage[letterpaper, left=0.7in, right=0.7in, bottom=0.99in, top=0.7in]{geometry}
\IEEEoverridecommandlockouts

\usepackage[font=footnotesize,caption=false]{subfig} 

\usepackage{cite}
\usepackage{amsmath,amssymb,amsfonts, mathtools, amsthm}
\usepackage{algorithmic}
\usepackage{graphicx}
\usepackage{textcomp}
\usepackage{xcolor}
\usepackage{booktabs}
\usepackage{glossaries}
\usepackage{subfig}
\usepackage{algorithm}
\usepackage{algorithmic}
\usepackage{bbm}
\usepackage{bm}
\usepackage{multirow}

\usepackage[normalem]{ulem}
\usepackage{verbatim}
\usepackage{fancyhdr}

\newacronym{bcd}{BCD}{Block Coordinate Descent}
\newacronym{leo}{LEO}{Low Earth Orbit}
\newacronym{isl}{ISL}{Inter-Satellite Link}
\newacronym{gsd}{GSD}{ground sample distance}
\newacronym{fov}{FoV}{field of view}
\newacronym{gtfp}{GTFP}{ground track frame period}
\newacronym{gs}{GS}{ground station}
\newacronym{fso}{FSO}{free-space optical}
\newacronym{smec}{SMEC}{satellite mobile edge computing}
\newacronym{mec}{MEC}{mobile edge computing}
\newacronym{pdd}{PDD}{Penalty Dual Decomposition}
\newacronym{aodv}{AODV}{Ad hoc On-demand Distance Vector}
\newacronym{cnn}{CNN}{convolutional neural networks}
\newacronym{dl}{DL}{Deep Learning}
\newacronym{dod}{DoD}{depth of discharge}
\newacronym{dqn}{DQN}{deep Q-learning}
\newacronym{gsl}{GSL}{ground-to-satellite link}
\newacronym{bm}{BM}{benchmark}
\newacronym{ml}{ML}{Machine Learning}
\newacronym{mdp}{MDP}{Markov decision process}
\newacronym{ngeo}{NGEO}{Non-geostationary orbit}
\newacronym{olsr}{OLSR}{optimized link state routing protocol}
\newacronym{ospf}{OSPF}{Open Shortest Path First}
\newacronym{pan}{PAN}{Path-Aware Networking}
\newacronym{qos}{QoS}{Quality of Service}
\newacronym{rl}{RL}{Reinforcement Learning}
\newacronym{drl}{DRL}{Deep \gls{rl}}
\newacronym{dnn}{DNN}{Deep Neural Network}
\newacronym{dql}{DQL}{Deep Q-learning}
\newacronym{e2e}{E2E}{end-to-end}
\newacronym{bgp}{BGP}{Border Gateway Protocol}
\newacronym{ibgp}{iBGP}{interior Border Gateway Protocol}
\newacronym{ebgp}{eBGP}{exterior Border Gateway Protocol}
\newacronym{as}{AS}{Autonomous System}
\newacronym{relu}{ReLu}{Rectified Linear Unit}
\newacronym{cdf}{CDF}{Cumulative Distribution Function}
\newacronym{ntn}{NTN}{Non-Terrestrial Networks}
\newacronym{lsatc}{LSatC}{\gls{leo} Satellite Constellation}
\newacronym{ai}{AI}{Artifical Intelligence}
\newacronym{ip}{IP}{Internet Protocol}
\newacronym{ue}{UE}{User Equipment}
\newacronym{pomdp}{POMDP}{Partially Observable Markov Decision Problem}
\newacronym{hol}{HOL}{Head of Line}
\newacronym{fifo}{FIFO}{First-In First-Out}
\newacronym{snr}{SNR}{Signal-to-Noise Ratio}
\newacronym{eo}{EO}{Earth Observation}
\newacronym{aoi}{AoI}{Age of Information}
\newacronym{paoi}{PAoI}{Peak Age of Information}
\newacronym{semcom}{SemCom}{Semantic Communications}
\newacronym{go}{GO}{Goal-Oriented Communications}

\def\BibTeX{{\rm B\kern-.05em{\sc i\kern-.025em b}\kern-.08em
    T\kern-.1667em\lower.7ex\hbox{E}\kern-.125emX}}

\title{Goal-oriented vessel detection with distributed computing in a \gls{leo} satellite constellation}

\author{\IEEEauthorblockN{Antonio Mercado-Martínez, Beatriz Soret~\IEEEmembership{Senior Member,~IEEE}, Antonio Jurado-Navas~\IEEEmembership{Member,~IEEE}}
\vspace{-0.4cm}

\thanks{This work is partially funded by ESA SatNEx V (prime contract no. 4000130962/20/NL/NL/FE), and by the Spanish Ministerio de Ciencia, Innovación y Universidades (PID2022-136269OB-I00). The view expressed herein can in no way be taken to reflect the official opinion of the European Space Agency. The author thankfully acknowledges the computer resources, technical expertise and assistance 
provided by the SCBI (Supercomputing and Bioinformatics) center of the University of Malaga.} }

\def\subparagraph{} 
\usepackage{titlesec}
\titlespacing*{\section}{0pt}{*1}{*1}
\titlespacing{\subsection}{0pt}{*1}{*1}

\renewcommand{\thesubsubsection}{\arabic{subsubsection}}

\titleformat{\subsubsection}[runin]{\itshape}{\thesubsubsection)}{1em}{}
\titlespacing*{\subsubsection}{\parindent}{0pt}{*1}

\fancypagestyle{firstpage}{
  \fancyhf{}

  \fancyfoot[C]{979-8-3503-8481-9/24/\$31.00 \copyright2024 IEEE}
}

\begin{document}

\bstctlcite{IEEEexample:BSTcontrol}

\maketitle
\thispagestyle{firstpage}
\begin{abstract}
\gls{eo} has traditionally involved the transmission of a large volume of raw data to map the Earth surface. This results in congestion to the satellite network and delays in the availability of the results, invalidating the approach for timing-sensitive applications. Instead, the computation resources at the satellites can be used as an edge layer for compressing the data and/or doing inferences. In this paper, we investigate satellite edge computing for vessel detection with a \gls{leo} satellite constellation. First, we distribute the computation and inference load among the neighbouring satellites of the one taking the images, based on the VHRShips data set and YOLOv8. This semantic and fragmented information is then routed to a remote ground monitor through the whole constellation. The average and peak \gls{aoi} are reformulated to measure the freshness of the aggregated information at the receiver in this image-capture scenario. We then dimension the network -- number of orbital planes and satellites per orbital plane -- for a given target age and covered area that quantify the level of achievement of the task. The results show that  $20$ orbital planes with $20$ satellites are necessary to keep the peak \gls{aoi} below \mbox{$60$ s} with a compression ratio $>23000$, i.e., a size reduction of $99.996\%$, and for a $\sim100\%$ probability of coverage. 

\end{abstract}

\glsresetall

\section{Introduction}

Advancements in \gls{ai} and \gls{dl}  have revolutionized communication networks and the way information is transmitted, bringing concepts such as \gls{semcom} and \gls{go} to the forefront\cite{denizjsac2023}. \gls{semcom} refers to the idea of efficiently transmitting the key elements of the message, rather than the message in its entirety. \gls{go} is an intimately related approach wherein the receiver is interested in the effectiveness of the source’s transmitted message to accomplish a certain goal. Ultimately, they can lead to significant network decongestion and lower transmission times, which is of paramount importance in space.

\begin{figure}
\centering
\includegraphics[width=0.35\textwidth]{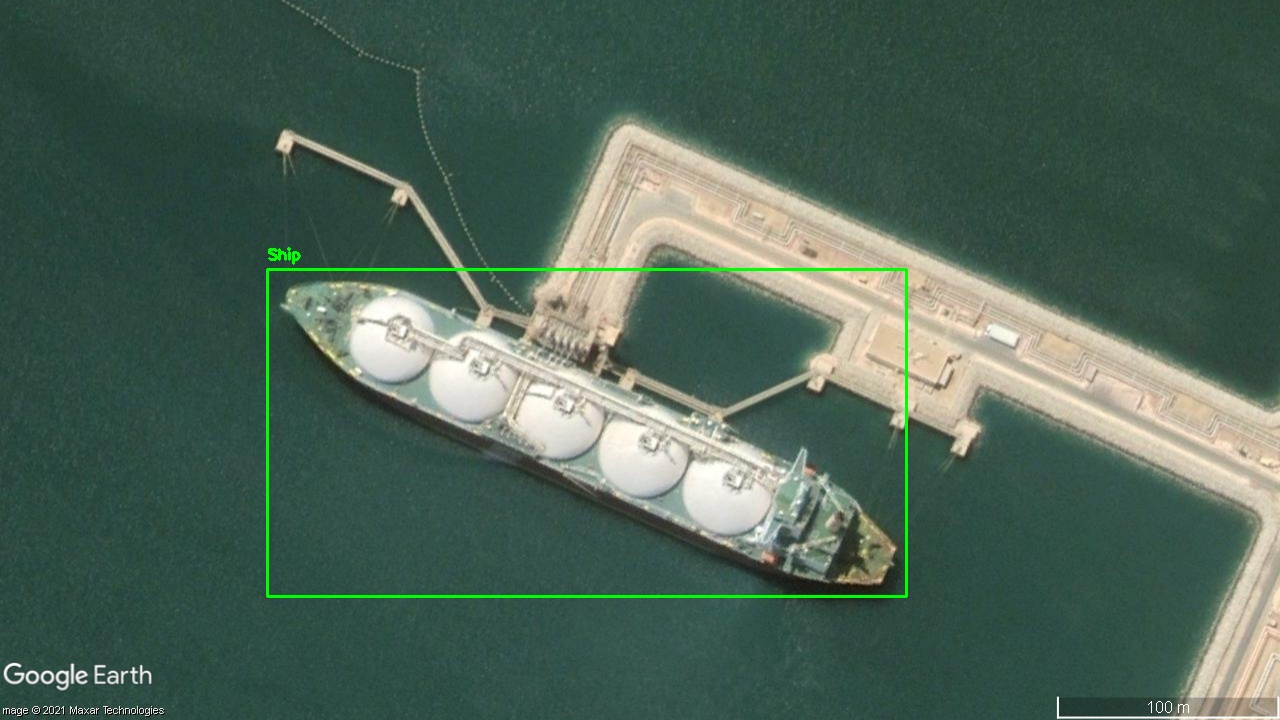}
\caption{Example of YOLOv8 vessel detection algorithm \cite{yolov8_ultralytics} performance over an image of the VHRShips dataset\cite{ijgi11080445}.}
\label{fig:yolov8_example} \vspace{-0.45cm}
\end{figure}

\gls{eo} is a historical application of satellites, yet, its importance is consistently increasing. This is due to its potential to provide unique information for climate and environmental monitoring, maritime surveillance, or disaster management, among others\cite{Leyva_Mayorga_2023}. The principles of \gls{semcom} and \gls{go} are very interesting for \gls{eo}, as the scenario involves handling a large amount of information and making inferences that depend on the system's goal. The covered area and the quality of the captured frames depends on various factors, but some of the most relevant are: (1) the \gls{fov} of the camera of the satellite, which is the angle that determines the observable space of the camera sensor; (2) the altitude ($h$) of the deployment, being \gls{leo} satellites a good choice in practice (e.g., Sentinel and Landsat) due to its ability to capture high-resolution images and the frequent revisit time; (3) and the \gls{gsd}, which is the real distance between the center of two adjacent pixels. A small value of \gls{gsd} improves the image definition. 

Within the landscape of \gls{eo} applications, we  investigate near real-time vessel detection using a constellation of \gls{leo} satellites, and leverage the distributed processing capacity of the satellites. The task -- taking the image, detecting the vessels, and transmitting the bounding boxes  (see Fig.~\ref{fig:yolov8_example}) -- is a complex task that involves a significant processing load. To enable the near real-time detection by reducing the time elapsed between capturing the frame and making the detection available at the remote \gls{gs}, the task is distributed in parallel among multiple neighbouring satellites, working as an edge computing layer, and then the accomplished detection is transmitted to the \gls{gs}.  The focus is on the trade-off between the detection accuracy and the time, and between the computation and communication resources. The framework is used to dimension the constellation with the requirement of achieving the highest probability of coverage of the mapped area while ensuring that the information reaching the ground is as fresh as possible. For the latter, we use \gls{aoi} as a suitable \gls{go} metric for detection and tracking applications. The rest of the paper is organized as follows. Section~\ref{sec:soa} provides an overview of related works. Section~\ref{sec:systemmodel} presents the system model, while the design is introduced in Section~\ref{sec:timingframework}. Results are discussed in Section~\ref{sec:results}, and Section~\ref{sec:conclusions} concludes the paper.

\section{State of the Art} \label{sec:soa}

Vessel detection has been extensively studied recently, so we have plenty of references on the subject. Datasets like Ships From Google Earth \cite{ships-google-earth_dataset} or VHRShips  \cite{ijgi11080445} collect remote sensing images to train \gls{ai} algorithms designed for detecting objects in images like YOLO with really good results ($\sim 90\%$ of accuracy) (Fig. \ref{fig:yolov8_example}). These datasets typically feature images with a small \gls{gsd} (of around 0.5 meters and at most 1.5 meters). Regarding the detection algorithms, they are computationally expensive compared to classical compression algorithms like JPEG, but the size of the resulting semantic information (the coordinates of the bounding box(es) of the detected vessels) is significantly smaller than compressed images.

Edge computing has recently been applied to \gls{leo} satellite constellations to reduce network congestion and latency. In \cite{Leyva_Mayorga_2023}, an edge computing framework for \gls{eo} that optimizes processing load distribution and compression parameters to minimize energy consumption and increase processed information is formulated. \cite{10061289} proposes a scheduling approach involving observation, relay, and computing satellites to maximize the number of completed observation tasks.

\gls{semcom} and \gls{go} have been widely studied in terrestrial networks. We refer the reader to \cite{denizjsac2023} for an overview. The case of moving objects in ground has analogies with our tracking problem~\cite{nikos2021}. In \gls{go} designs, the \gls{aoi} is a common used metric to measure the elapsed time since the generation of the last received update, or, equivalently, the freshness of the information \cite{8187436}. The age increases linearly in time if there are no updates, and it decreases upon reception of a new update to the time elapsed since such update was generated. The abundance of papers regarding the \gls{aoi} is overwhelming and there are many proposed variants, being the \gls{paoi} \cite{9808166} a good option for worst-case optimizations. 

\section{System model} \label{sec:systemmodel}
\begin{figure} [t]
\centering
\includegraphics[width=0.38\textwidth]{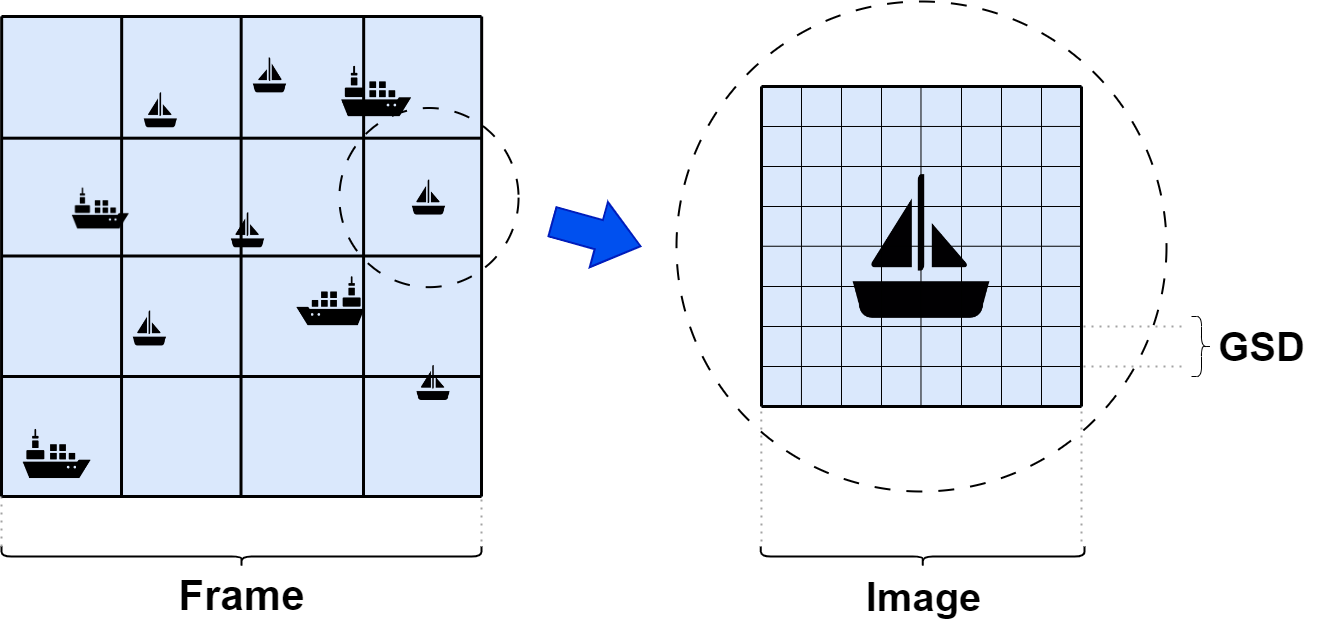}
\caption{Example of a frame captured and divided into images of a specific resolution.}
\label{fig:frame} \vspace{-0.4cm}
\end{figure}

The \gls{leo} satellite constellation consists of $M$ circular orbital planes deployed at an altitude $h$ and with a given inclination $\delta$. There are $N$ evenly distributed satellites in each of such orbital planes, i.e., the constellation size is $M$x$N$. These four parameters, $M$, $N$, $h$, and $\delta$, are the parameters of interest for the design and dimensioning of the constellation. These have to be selected taking into account performance parameters such as the availability of the service, i.e., the probability of being under the coverage area for a specific service \cite{2022}, which is also impacted by the latitude. Furthermore, the organization of satellites along the orbital shell (a set of orbital planes displayed at the same altitude) can be of two types: walker star or walker delta. A walker star orbital shell consists of nearly-polar orbits ($\delta$ $\sim 90$º) evenly spaced within 180º (the angle between adjacent orbital planes is 180/N), while a walker delta orbital shell consists of inclined orbits ($\delta$ $< 60$º) evenly spaced within 360º (the angle between adjacent orbital planes is 360/N).
\begin{figure}[t]
\centering
\includegraphics[width=0.38\textwidth]{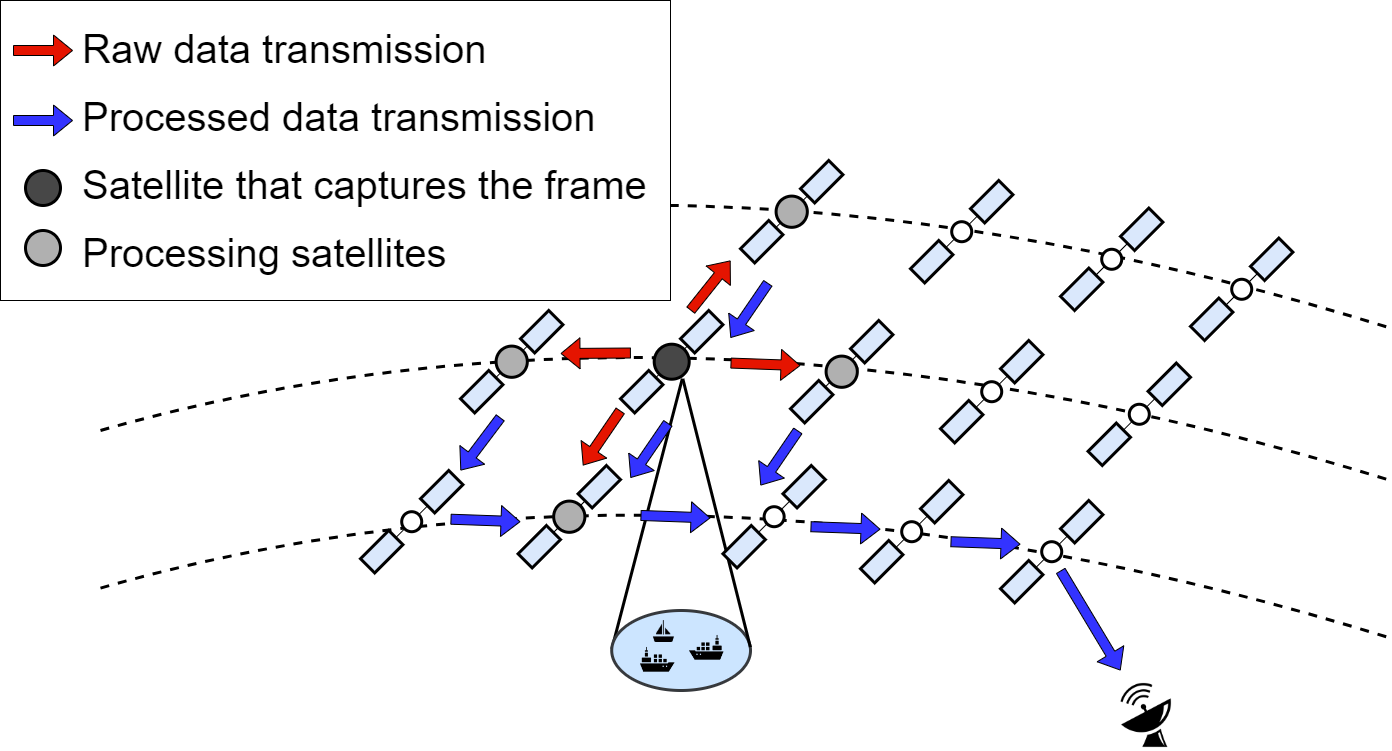}
\caption{Sketch of the scenario.} \vspace{-0.4cm}
\label{fig:scheme}
\end{figure}
The constellation is used to capture the images, process them, and send up-to-date information from a predetermined area to ground, i.e., to transmit the identified and located vessels. For this, the satellites are equipped with a communication payload and a processing payload. This information is monitored in a predefined \gls{gs}. The process is as follows (see Fig.~\ref{fig:scheme}): 
\begin{enumerate}
\item The area to be monitored is predefined.
\item A frame is taken when any of the satellites in the constellation is in a suitable position to capture a quality frame of the area. As this depends on various complex and interrelated factors, it will be considered that a quality photo can be taken up to a specific angle $\beta$ off-nadir. It will also be assumed that the atmospheric and meteorological conditions are suitable.
\item A vessel detection algorithm is applied in a distributed and parallel way by fragmenting the original frame. It is assumed that all satellites in the constellation will have the YOLOv8 \cite{yolov8_ultralytics} vessel detection algorithm on their CPUs, pre-trained with the Ships From Google Earth dataset \cite{ships-google-earth_dataset}, and will be capable of capturing images with sufficient quality for the algorithm to perform effectively. This algorithm has been selected because, compared to similar ones, it is computationally less expensive. However, the computation load is still considerable, as the amount of information to be processed, so it will be fragmented and equally divided among the satellite that takes the frame and its four neighbors (two in the same orbital plane and two in the adjacent orbital planes). The sections into which the frame has been divided are further divided into images of a specific resolution to be processed (Fig \ref{fig:frame}). 
\item The result of the detection is routed to a specific \gls{gs}. 
Once processing is complete, the fragmented information consisting of the detected vessels in each fragment is routed to the \gls{gs} by each satellite participating in the computing, and using the routing algorithm developed in \cite{article}. This derives the path with the shortest propagation distance between two satellites from the paths with the fewest number of hops between them (the minimum hop path region) based on the relationship between satellite phase and inter-satellite link distance. 
\end{enumerate}

It should be noted that before taking the next frame, the processing of the last one taken must be fully completed so that the processing load does not accumulate.

To calculate the amount of information sent to the \gls{gs}, it must be taken into account the compression factor ($\rho$) and the portion of images in the captured frame with detected vessels ($\alpha_{vessels}$). The compression factor is given by
\begin{equation}
    \rho=\frac{{D}_{img}}{N_{vessels}{D_{bbox}}}, 
\end{equation}
where ${D}_{img}$ is the average size of the image, $N_{vessels}$ the average number of detected vessels per image and $D_{bbox}$ the average size of the bounding boxes. In this way, the amount of information $x_{g}$ sent to the GS can be calculated as follows:
\begin{equation}
    x_{g} =\frac{x \alpha_{vessels}}{\rho},
\end{equation}
where $x$ is the amount of information to process in bits.

Regarding communications, \gls{fso} links will be used for intra-plane \gls{isl} and radio frequency (RF) links for inter-plane \gls{isl} and the downlink. The transmission rates for the links are obtained as described in \cite{2022}.

In addition, there are losses both in the communication (packet losses) and in the detection (vessels present in the images but not detected), and they both will affect the timing performance.
We will consider a packet loss probability on a link ranging from $P_{min}$ to $P_{max}$ depending on the distance between satellites. This will be given by the following formula:
\begin{equation}
    P_{loss}= P_{max} + (P_{min} - P_{max})\exp{\left(-\frac{d}{d_{max} - d_{min}}\right)},
\end{equation}
where $d$ the distance between the two satellites where the links, and $d_{max}$ and $d_{min}$ are maximum and minimum distances between adjacent satellites selected accordingly to the simulated constellations as explained in Section~\ref{sec:results}.  
For the computation and communication losses the worst-case scenario will be considered. In communications, losing one packet means that the whole frame is lost.  Likewise, the losses due to the detection algorithm will be determined by the recall, which relates the number of detected vessels to the actual number of vessels, obtained when testing the algorithm with the test dataset:
\begin{equation}
    \text{Recall} = \frac{\text{Number of detected vessels}}{\text{Actual number of vessels}}.
\end{equation}

\section{Constellation design and timing metrics} \label{sec:timingframework}
\subsection{Constellation design}
There are three elements in the system model related to the constellation design. The first one is to decide the value of some features of the constellation that make feasible to capture images with sufficient quality to perform the vessel detection process. Due to its homogeneity, the images VHRShips dataset, with a \gls{gsd} of 0.43 m/pixel, has been taken as reference. Moreover, the parameters of the satellite WorldView-3 \cite{worldview-3} are suitable to capture images comparable to those in the selected dataset, and therefore has been selected as a reference for the study. The altitude is approximately $600$ km. The second step is to the decide if the constellation consist of a walker star or a walker delta orbital shell. The criterion here is to cover the largest possible area, but also to ensure that there are the maximum number of satellites in areas of interest, i.e., most of the globe is covered most of the time. Taking into account that the vast majority of vessels located outside of the polar regions \cite{marine_traffic}, we choose a walker delta orbital shell with a sufficient number of satellites in regions of interest. Finally, the density of the constellation must be selected, i.e., the number of orbital planes $M$ and satellites per orbital plane $N$, for which we aim at ensuring that the average time between capturing one frame and the next one is below a predetermined threshold. These values will be obtained through simulations relying on the coverage probability ($P_m$) and the \gls{aoi} as defined next.

\subsection{Timing metrics}
Both communication and processing times contribute to the \gls{aoi}. The frame acquisition time is negligible. For the processing times, the execution times of the algorithm were measured on a personal computer, and these were converted into CPU cycles per bit. From a set of these measurements, a gamma function can be derived to randomize the complexity of the algorithm within a realistic range of values.
Thus, given the computational capacity of the satellites (in terms of the number of cores and CPU clock frequency), the processing times can be calculated. Processing times are calculated as:
\begin{equation}
    T_{proc}=\frac{xC}{N_{cores}f_{CPU}},
\end{equation}
where $C$ is the complexity of the algorithm in CPU cycles per bit, $N_{cores}$ the number of cores of the CPU and $f_{CPU}$ the CPU frequency.
Regarding communication, transmission and propagation times, they will be calculated from the satellite taking the frame to its neighbors, plus the transmission and propagation times from the processing-satellites to the \gls{gs}. For the transmission this is written as the sum over the links $l$, i.e., 
\begin{equation}
    T_{trxon}=\sum_{l}\frac{D_{l}}{R_{l}},
\end{equation}
where $D_{l}$ is the number of bits to transmit through a given link $l$ and $R_{l}$ the bit rate of such link. And for the propagation:
\begin{equation}
    T_{prop}=\frac{\sum_l{d_{l}}}{c},
\end{equation}
where $d_{l}$ is the distance between two nodes of the path obtained from the routing algorithm \cite{article}, which determines the path with the minimum propagation distance between two satellites among those with the minimum number of hops; and $c$ the speed of light. 

\begin{figure} [t]
\centering
\includegraphics[width=0.36\textwidth]{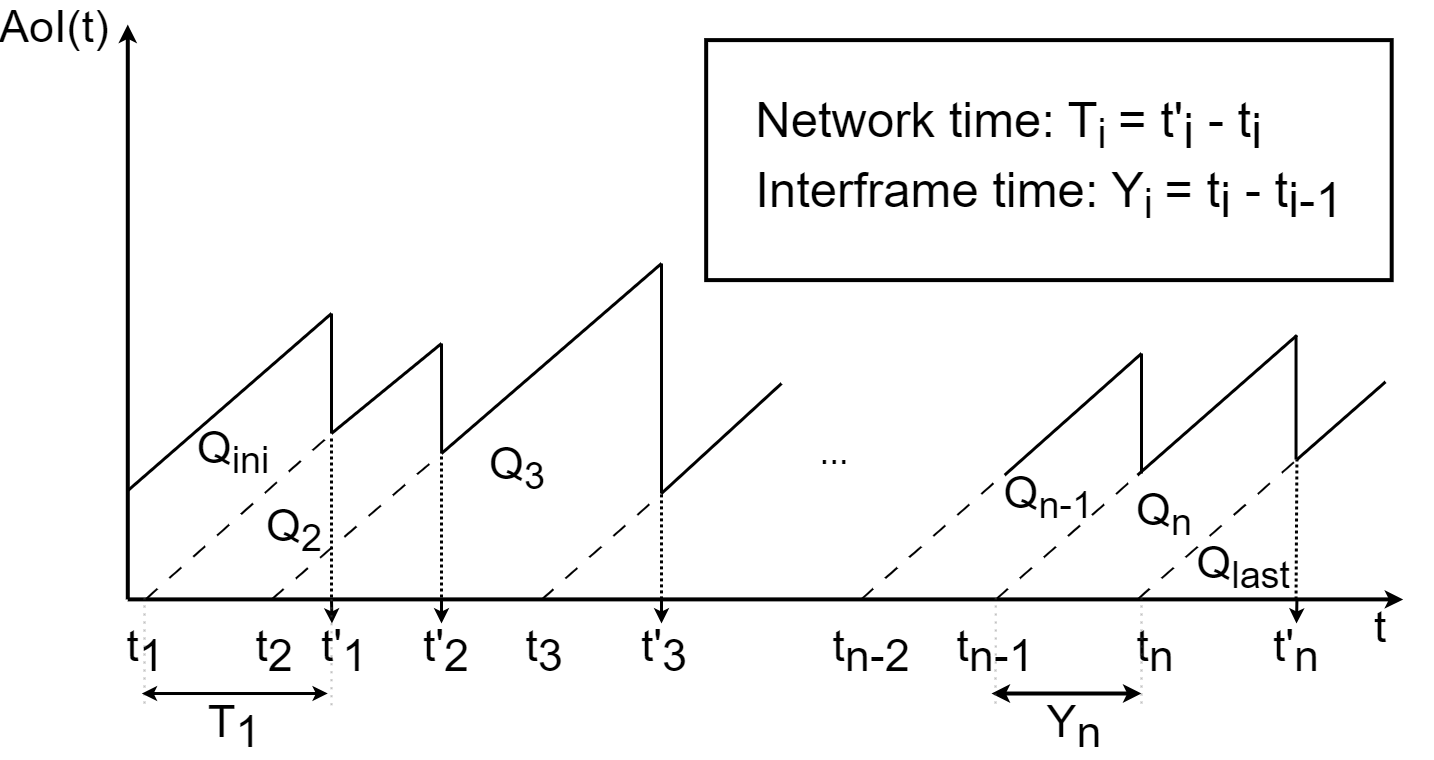}
\caption{Evolution of the \gls{aoi}.}
\label{fig:AoI} \vspace{-0.4cm}
\end{figure}

The \gls{aoi} is reformulated to consider the instant at which the frames are captured and at which the information of the mapped area reaches the \gls{gs}.
Mathematically, this is expressed as follows. 
Fig. \ref{fig:AoI} shows the time evolution of the \gls{aoi}. It is assumed that the system is first observed at $t = 0$ and index $i$ represents the number of the frame. A frame $i$ is captured at $t = t_i$ and its semantic information is fully received at the \gls{gs} at $t = t'_i$. As the processing task is distributed in parallel among multiple satellites and the resulting data is routed through different paths to the destination, there will be several timestamps representing the reception of semantic information in the \gls{gs}, of which the largest one will be considered, representing the time in which all processed information is available at the \gls{gs} and therefore the remote monitor has the full information of the mapped area. Assuming the processing is distributed among $n$ satellites:
\begin{equation}
    t'_i=\text{max}\left\{{t'}_i^{(1)}, {t'}_i^{(2)}, \text{...}, {t'}_i^{(n - 1)}, {t'}_i^{(n)}\right\}.
\end{equation}
$T_i$ is defined as the total network time of the system (communication + processing times); and $Y_i$ as the interframe time $T_i = t'_i - t_i$, the time between the capture of the frames whose semantic information is subsequently received $Y_i = t_i - t_{i - 1}$.

The average \gls{aoi} can be calculated as follows:

\begin{equation}
    \text{AoI}_{avg}=\frac{1}{\tau}\left(Q_{ini} + Q_{last} + \sum_{i = 2}^{N(\tau)}{Q_i}\right);
\end{equation}
where $\tau$ is the total observation time and $N(\tau)$ the the number of arrivals by that time. As can be seen in Fig. \ref{fig:AoI}, $Q_i$ for $1 < i$ are trapezoids whose areas can be calculated as:

\begin{equation}
    Q_i=\frac{1}{2}\left(T_i + Y_i\right)^2 - \frac{1}{2}T_i^2 = Y_iT_i + \frac{Y_i^2}{2}.
\end{equation}
Likewise, assuming that the system is observed for the first time at $t = 0$, the average \gls{paoi} can be calculated as:

\begin{equation}
    \text{PAoI}_{avg}=\frac{1}{N(\tau)}\left(t'_1 + \sum_{i = 2}^{N(\tau)}{(t'_i - t_{i - 1})}\right).
\end{equation}

\section{Simulation and results} \label{sec:results}

The first orbital shell of Starlink has been used as the reference constellation because its altitude ($\sim 550$ km) is suitable for obtaining quality images and the orbital shell is of the walker delta type ($\delta = 53$º) . In order to obtain the density of the constellation which allows to have an average \gls{paoi} value below a certain threshold, two series of simulations will be carried out. The first one will be conducted by fixing $N$ and varying $M$ to determine the number of orbital planes and, once it has been determined, the second one will be conducted by fixing $M$ and varying $N$ to determine the number of satellites per orbital plane. 
Regarding the packet loss probability, values between $P_{min} = 0.001$ and $P_{max} = 0.1$ have been considered, as they can be considered typical values \cite{p_loss}, and $d_{max}$ and $d_{min}$ are the maximum and minimum distances between satellites for the constellations considered for the simulations, respectively. For the computation losses, a recall of 0.9 is assumed. It will be considered that the information is updated with sufficient frequency if the average \gls{paoi} is less than 60 seconds.

\begin{figure}
\centering
\includegraphics[width=0.4\textwidth]{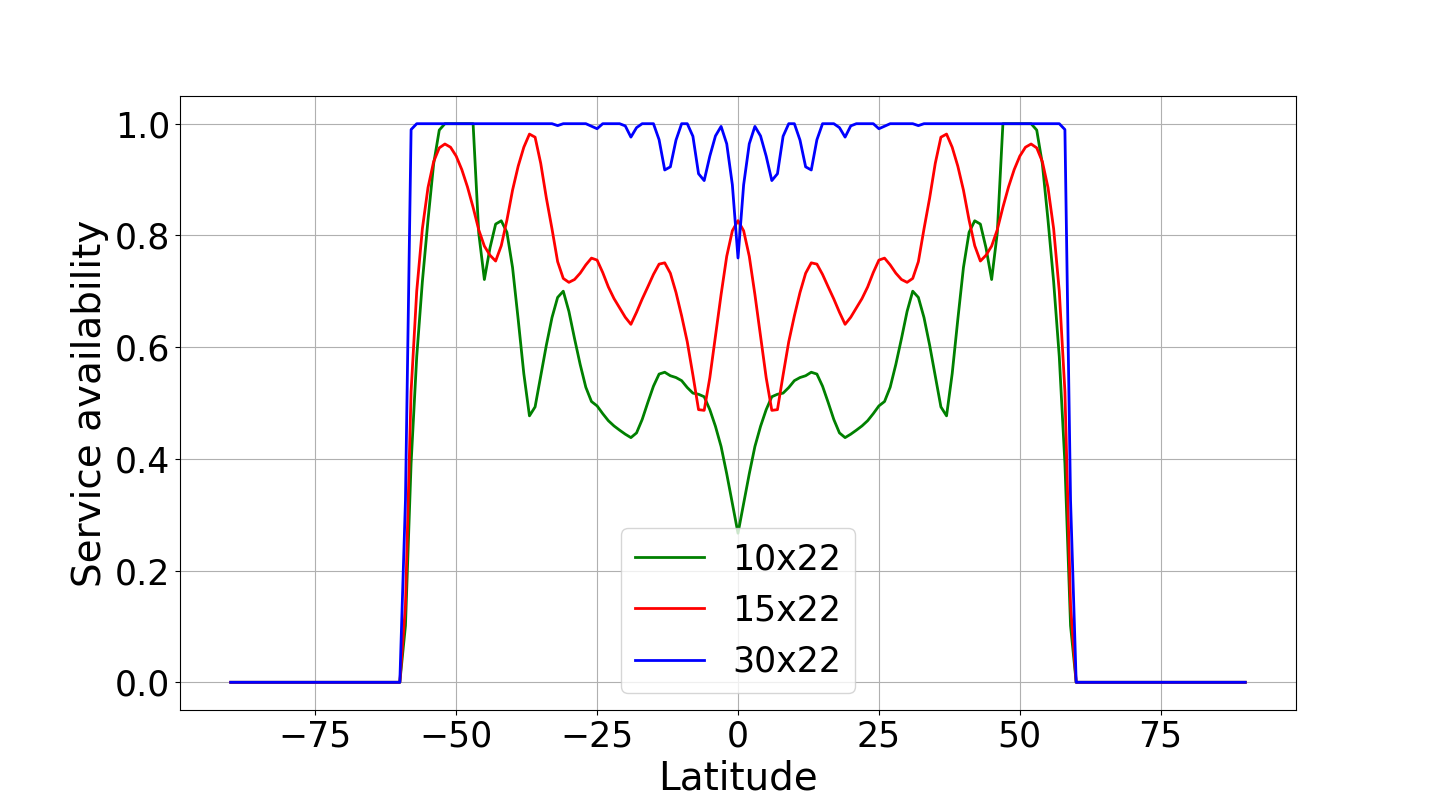}
\caption{Service availability taking a quality frame for vessel detection. Starlink-like topology with $M = 10, 15, 30$, and $N = 22$.}
\label{fig:cov_prob} \vspace{-0.4cm}
\end{figure}

\begin{figure}
\centering
\includegraphics[width=0.35\textwidth]{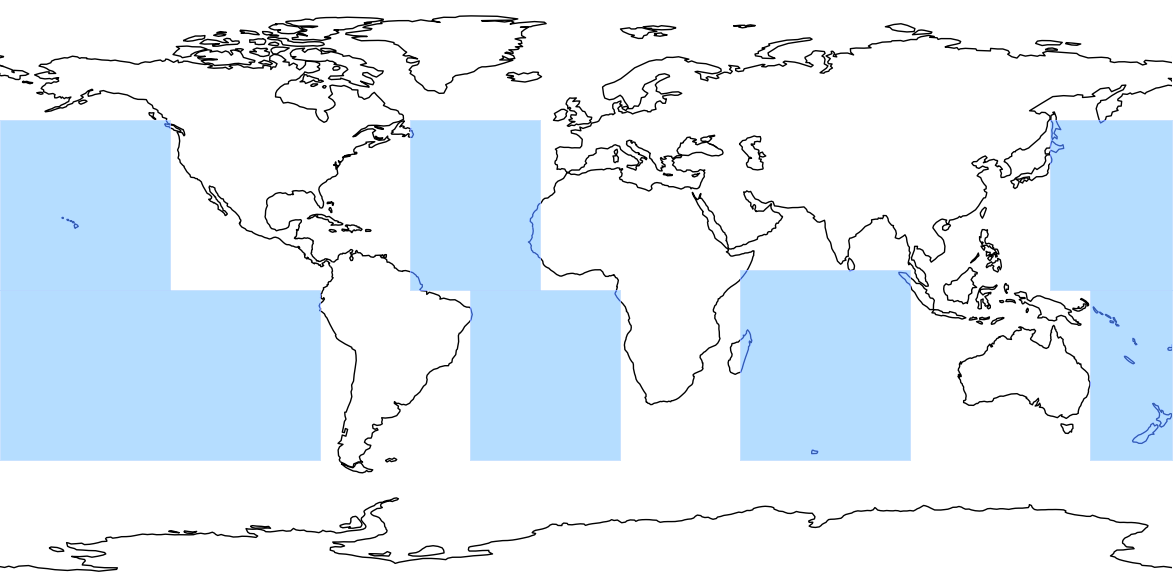}
\caption{Coordinates prone to be selected for water-covered regions.}
\label{fig:map} \vspace{-0.4cm}
\end{figure}

The simulations aim to calculate the \gls{aoi} over a period of Earth rotation to obtain the average \gls{aoi} and \gls{paoi} for a specific coordinate to be monitored. As the probability of taking a quality frame depends on the coordinate to be monitored (see Fig. \ref{fig:cov_prob}) a series of Monte Carlo simulations will be conducted by randomizing it, ultimately obtaining the average values. To ensure that the area to be monitored is in a water-covered region, the coordinates prone to be selected have been restricted to those shown in Fig. \ref{fig:map} and the \gls{gs} to which the resulting information will be sent will be located in Los Angeles (latitude 34.05º, longitude -118.24º). It will be assumed that a certain amount of vessels are always in the area to be monitored and that the area covered by a frame is 162.16 km². 
Other relevant parameters for the simulation are: $f_{CPU} = 1.8$ kHz, $N_{cores} = 8$, $\beta = 50$º, $D_{img} = 391.43$ kB, \gls{gsd} = 0.43 m/pixel, an image resolution of 720x1280 pixels, $D_{bbox} = 67.2$ bits, an average $C$ of 374.2 CPU cycles per bit, $\rho = 23299.4$, and $\alpha_{vessels} = 0.2$.

\begin{figure}[t]
\centering
\includegraphics[width=0.39\textwidth]{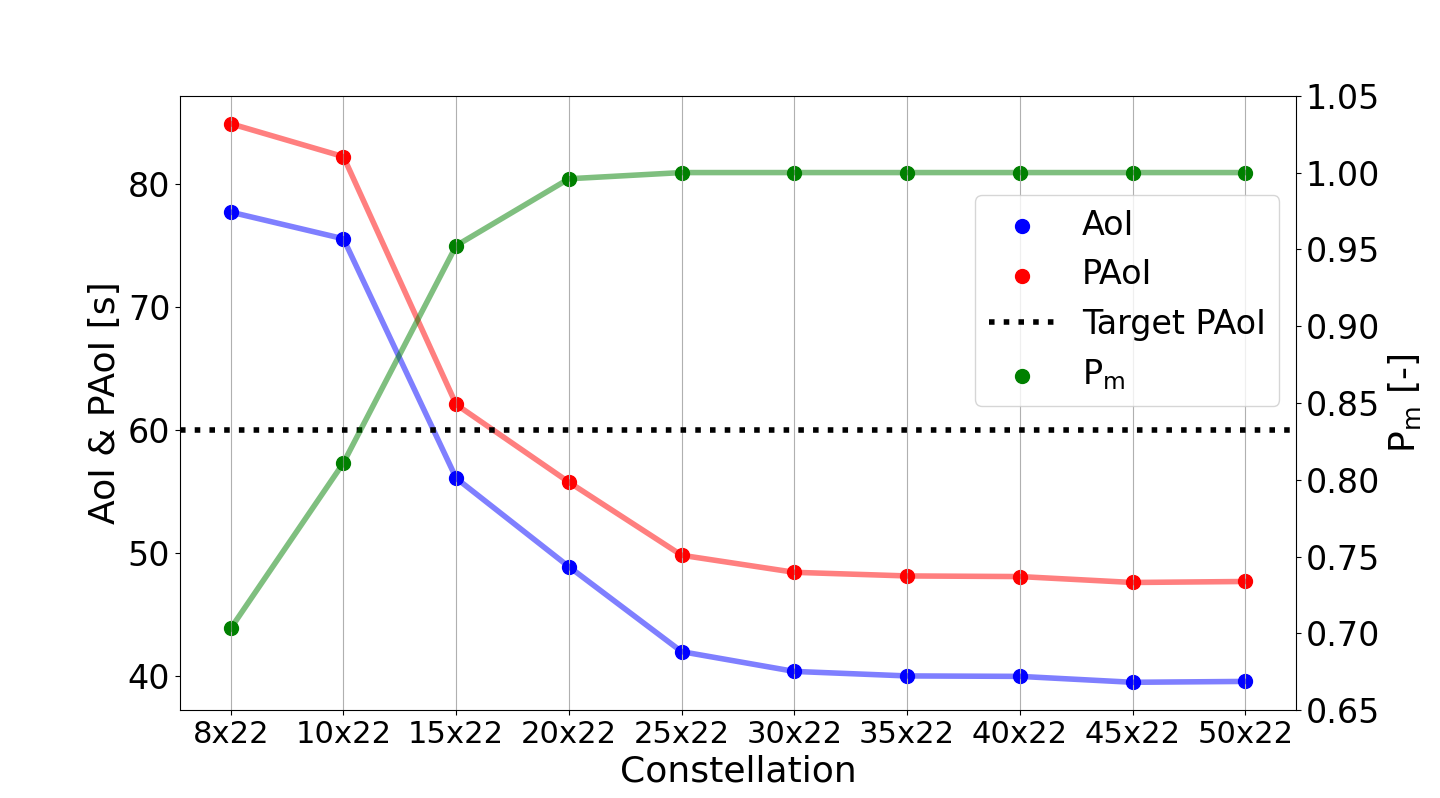}
\caption{Average \gls{aoi} and \gls{paoi} and probability of coverage of the monitored area ($P_m$) for Starlink-like topology with $N=22$ and varying $M$.}
\label{fig:results_AoI_N_fixed} \vspace{-0.4cm}
\end{figure}

\begin{figure}[t]
\centering
\includegraphics[width=0.39\textwidth]{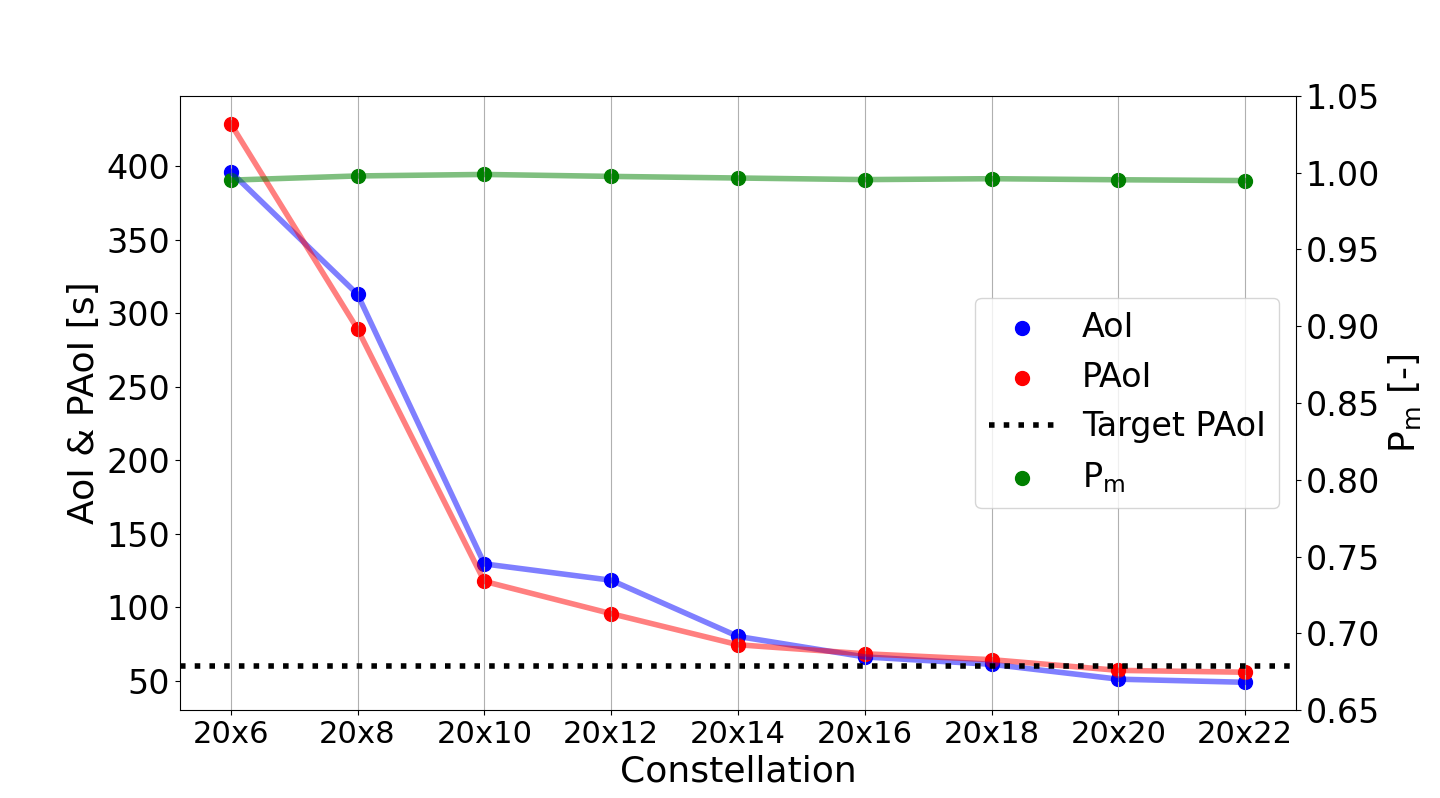}
\caption{Average \gls{aoi} and \gls{paoi} and probability of coverage of the monitored area ($P_m$) for Starlink-like topology with $M=20$ and varying $N$.}
\label{fig:results_AoI_M_fixed} \vspace{-0.4cm}
\end{figure}

Fig. \ref{fig:results_AoI_N_fixed} shows the results of the first sweep, keeping $N$ fixed, displaying the average \gls{aoi} and \gls{paoi}, along with the probability $P_m$ of covering the specific monitored area over a rotation period of the Earth for different number of orbital planes. It concludes that 20 orbital planes ensure an average \gls{paoi} below the established threshold with $P_m \sim 100\%$. Thus, the results of the second sweep, with $M = 20$, are shown in Fig. \ref{fig:results_AoI_M_fixed}, also displaying the average \gls{aoi} and \gls{paoi} along with $P_m$. According to them, it is possible to achieve an average \gls{paoi} below the threshold with $N = 20$. $P_m$ does not vary significantly.

\begin{figure}[t]
\centering
\includegraphics[width=0.39\textwidth]{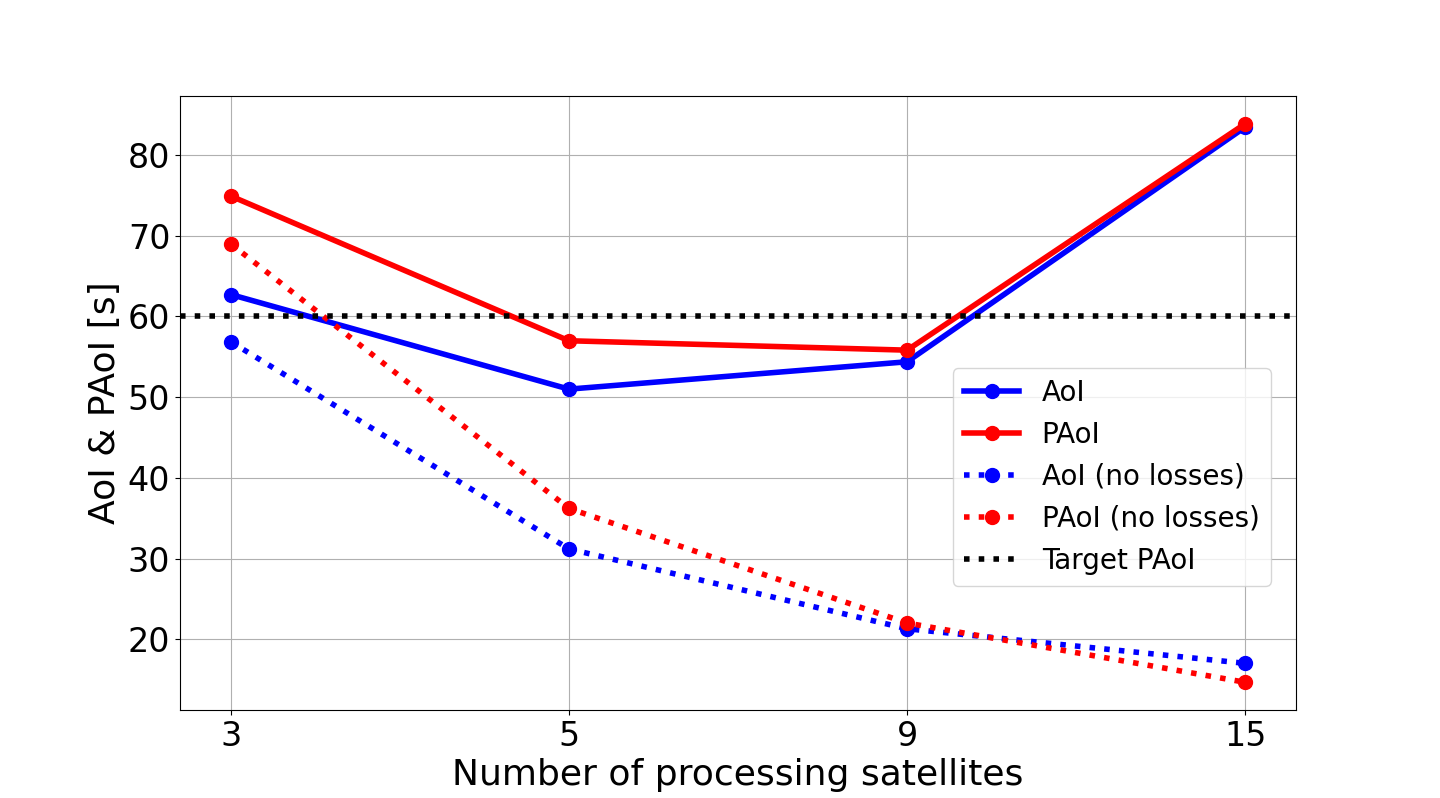}
\caption{Average \gls{aoi} and \gls{paoi} for a Starlink-like topology with $M=20$ and $N=20$ varying the number of processing satellites.
}
\label{fig:results_AoI_n_proc_sats} \vspace{-0.4cm}
\end{figure}

Once the constellation size is determined, the effect of varying the number of processing satellites is assessed. Fig.~ \ref{fig:results_AoI_n_proc_sats} shows the average \gls{aoi} and \gls{paoi} with different numbers of processing satellites. Without losses, more processing satellites reduce both \gls{aoi} and \gls{paoi}. However, considering losses, five processing satellites (the one taking the frame and its four neighbors) is adequate for the proposed scenario. Fewer processing satellites fail to meet the target \gls{paoi}, and more of them do not improve results and may worsen them due to increased packet loss probability.

Thus, five cooperating satellites in a walker delta orbital shell with $h = 550$ km, $\delta = 53$º and a size of 20x20 can accomplish the task while keeping the \gls{paoi} below 60 seconds. Furthermore, the amount of information to be transmitted is reduced from 2.98 Gb (the average frame size) to \mbox{127.81 Kb} (the total size of the bounding boxes for the simulation parameters), i.e., a size reduction of 99.996\%.

\section{Conclusions} \label{sec:conclusions}
In this paper, we investigate the potential of a \gls{leo} satellite constellation for near real-time vessel detection in a predefined geographical area, leveraging the distributed computation and communication capabilities of the spacecrafts. We apply a real data set and state-of-the-art YOLOv8 algorithm for the detection, and dimension the constellation for the best tradeoff between freshness of the information at the receiver and accuracy in the accomplishment of the task. The presented model can serve as a complement to Automatic Identification Systems (AIS) for vessel tracking \cite{marine_traffic}. Future work will generalize the model by considering the queries from ground (i.e., a closed-loop communication), the impact of atmospheric turbulences, and more comprehensive simulations.

\bibliographystyle{IEEEtran}
\bibliography{refs}

\end{document}